\documentclass[aps,prl,twocolumn,superscriptaddress,showpacs,showkeys]{revtex4-1}
\usepackage[english]{babel} 
\usepackage{tikz}
\usepackage{bm}       
\usepackage{amssymb}  
\usepackage{amsmath}
\usepackage[utf8]{inputenc}
\usepackage{mathrsfs}
\usepackage{hyperref}
\usepackage{color}
\usepackage[normalem]{ulem} 
\usepackage{braket}

\begin{document}

\title{Solitary wave propagation in media with step-like inhomogeneities}

\author{Mariya Lizunova}
\affiliation{Institute for Theoretical Physics, Utrecht University, Princetonplein 5, 3584 CC Utrecht, The Netherlands}
\affiliation{Institute for Theoretical Physics Amsterdam, University of Amsterdam, Science Park 904, 1098 XH Amsterdam, The Netherlands}

\author{Oleksandr Gamayun}
\email{o.gamayun@uva.nl}
\affiliation{Institute for Theoretical Physics Amsterdam, University of Amsterdam, Science Park 904, 1098 XH Amsterdam, The Netherlands}
\affiliation{Bogolyubov Institute for Theoretical Physics
14-b Metrolohichna str. Kyiv, 03143, Ukraine}

\date{\today}

\begin{abstract}

We consider the wave propagation in media with step-like inhomogeneities. We choose two different protocols: (I) a step-like spatial dependence of the coupling constant that physically corresponds to the junction of two systems and (II) the step-like time-dependence of the coupling constant that corresponds to a quench protocol. In both scenarios, we study the propagation of the solitary wave in $\varphi^4$ and the sine-Gordon model.  Due to topological protection, the solitary wave retains its form. This allows us to give model-independent predictions about its final velocity, which can be interpreted as the ``hyperbolic'' Snell's law. 
\end{abstract}


\maketitle 

Solitary waves are remarkable phenomena of non-linear physics. Their investigation started back in 1834 with the famous J.S. Russell's observation of the wave traveling for miles along the Union Canal near Edinburgh, Scotland, without altering its shape or speed~\cite{russel}. The theory of integrable non-linear equations was developed as an attempt to understand this phenomenon
\cite{Ablowitz_1981,Faddeev_1987}.  
The persistence of the wave is attributed to the existence of the infinite number of conserved quantities. Another remarkable result is the existence of two and more solitary waves with elastic scattering proprieties. These waves were dubbed solitons. They are now known to occur and play an important role in almost all areas of physics, including particle physics~\cite{skyrme}, cosmology~\cite{book3,friedland}, (non-linear) optics~\cite{chen,book2}, condensed-matter theory~\cite{bishop,strecker,katnelson}, Josephson-junction arrays~\cite{pendula,fillipov}, magnetic materials~\cite{maki}, convecting nematic fluids~\cite{lowe}, biophysics~\cite{brizhik}, and the analysis of seismic data~\cite{bykov}.

Another way to get a solitary wave is to consider topologically protected solutions of the non-linear equation \cite{manton}.  These waves are termed either topological solitons, or more specifically, depending on the model as kinks, skyrmyons, instantons, and so on \cite{radjaraman01}. Contrary to the integrable solitons, these waves can be found in higher dimensions, however, exact multisoliton solutions are usually impossible. 

In reality, integrable systems are always perturbed, but for small perturbations, the soliton maintains its integrity to some degree \cite{RevModPhys.61.763}.  
A sudden large perturbation of the coupling constant (a quench) may result in the splitting of the solitons even if the system remains integrable \cite{Gamayun_2016,PhysRevA.91.031605,PhysRevA.99.023605,Franchini_2015,Caudrelier_2016,10.1143/PTPS.55.284,Miles_1981}. Similar results are expected for solitons propagating in inhomogeneous media. Indeed, due to the localized form of a soliton and its relatively high velocity, one can expect that the soliton would ``self-quench'' while traveling between regions with different coupling constants. For the Korteweg-de Vries (KdV) waves this results in the gradient-induced fission of solitons \cite{PhysRevLett.27.1774,doi:10.1080/03091920701640115,Pelinovsky_2010}. 
For the non-linear Schrodinger (NLS) solitons, similar behavior of solitons was predicted for interfaces in optical fibers \cite{Abdullaev_1991,PhysRevE.53.2823}
and for beams of light reflected from and transmitted to the dielectric media 
\cite{Aceves:88,PhysRevA.39.1809,PhysRevA.39.1828,PhysRevA.41.1677}. 

We consider a similar setup for the topologically protected solitons that are present in the $\varphi^4$ model and the sine-Gordon model. We demonstrate that both for the quench protocol and the propagation in the inhomogeneous media the particle-like nature of the solitary wave in unaltered even in the highly non-perturbative regime. This allows us to give a quantitative prediction for the wave dynamics.

{\emph{Model}.} We consider a classical, real, scalar field $\varphi(t,x)$ in $(1+1)-$dimensional space-time with a step-like inhomogeneity of the coupling constant. The corresponding Lagrangian reads as 
\begin{equation}\label{eq:ener_lagrangian}
	\mathscr{L}=\frac{1}{2} \varphi_t^2-\frac{1}{2} \varphi_x^2-\Theta(t,x)\, U(\varphi).
\end{equation}
Here and below $\varphi_x = \partial \varphi/\partial x$ and $\varphi_t = \partial \varphi/\partial t$. 
We study two different types of inhomogeneity. In case I, we put
\begin{equation}\label{eq:theta1}
\Theta^{I}(t,x) =\zeta_1\theta (-x)+\zeta_2\theta(x),\enskip-\infty<t<\infty,
\end{equation}
where $\theta$ denotes the Heaviside step function. Physically, this corresponds to two semi-infinite intervals, with different prefactors $\zeta_1$ and $\zeta_2$, forming a junction at the point $x=0$. In case II,
\begin{equation}\label{eq:theta2}
\Theta^{II}(t,x) =\zeta_1\theta(-t)+\zeta_2\theta(t),\enskip-\infty<x<\infty.
\end{equation}
This corresponds to a quench of the coupling constant at the moment $t=0$ from $\zeta_1$ to $\zeta_2$, homogeneously in the entire space. We consider $\zeta_1,\,\zeta_2\ge 0$ to remain in the class of non-negative potentials $U(\varphi)$.

We choose the sine-Gordon and the $\varphi^4$ models with the potentials $U_{sG}=1-\cos\varphi$ and $U_{\varphi^4}=(1-\varphi^2)^2/4$, respectively. The first model is integrable resulting in the elastic scattering of the solitons. For the second one
dynamics of kinks is much more complicated and can be determined only approximately~\cite{campbell}.

As the initial state, we consider a solitary wave that is a solution of the equation of motions that corresponds to Eq.~\eqref{eq:ener_lagrangian}, for the homogeneous case $\zeta_2= \zeta_1$. It has a form 
\begin{equation}\label{eq:sol_generalform}
\varphi^{(0)}(x,t) = F\left(\sqrt{\zeta_{1}}\, \frac{x-x_0 -v_{in}\,t}{\sqrt{1-v_{in}^2}}\right),
\end{equation}
where $x_0$ is the initial position and ${0\le v^2_{in}<1}$ is the initial velocity of the solitary wave. The exact profile has a form of soliton for the sine-Gordon model  ${F_{sG}(x)=4\,\text{arctan}\left( \text{exp} (x)\right)}$ and the kink ${F_{\varphi^4}(x)=\tanh\left(x/\sqrt{2}\right)}$ for the $\varphi^4$ model.

For both models, we numerically study the evolution of the solitary wave using a method of finite-difference for open boundary conditions~\cite{Courant} with the continuity equation verification on each time step. To characterize the wave evolution we calculate the final velocity $v_f$ by  studying the position of the center as a function of $t$ to obtain $v_f$. A systematic error for $v_f$ is less than $2.2\%$, so we do not put error-bars on the plots below. 

{\it Junction.} We start our consideration with the spatially inhomogeneous case $I$ (see Eq.~\eqref{eq:theta1}). We initialize the moving soliton/kink far from the boundary $x_0 \ll -1/\sqrt{\zeta_1}$ with the positive initial velocity $v_{in}$. Qualitatively we observe that depending on the values of $v_{in}$ and $\zeta_2$ the wave is either elastically reflected from the boundary or completely transmitted. In all considered cases the solitary wave keeps its form and the radiation modes have very small amplitude.

\begin{figure}[h]
\center{\includegraphics[width=0.9\linewidth]{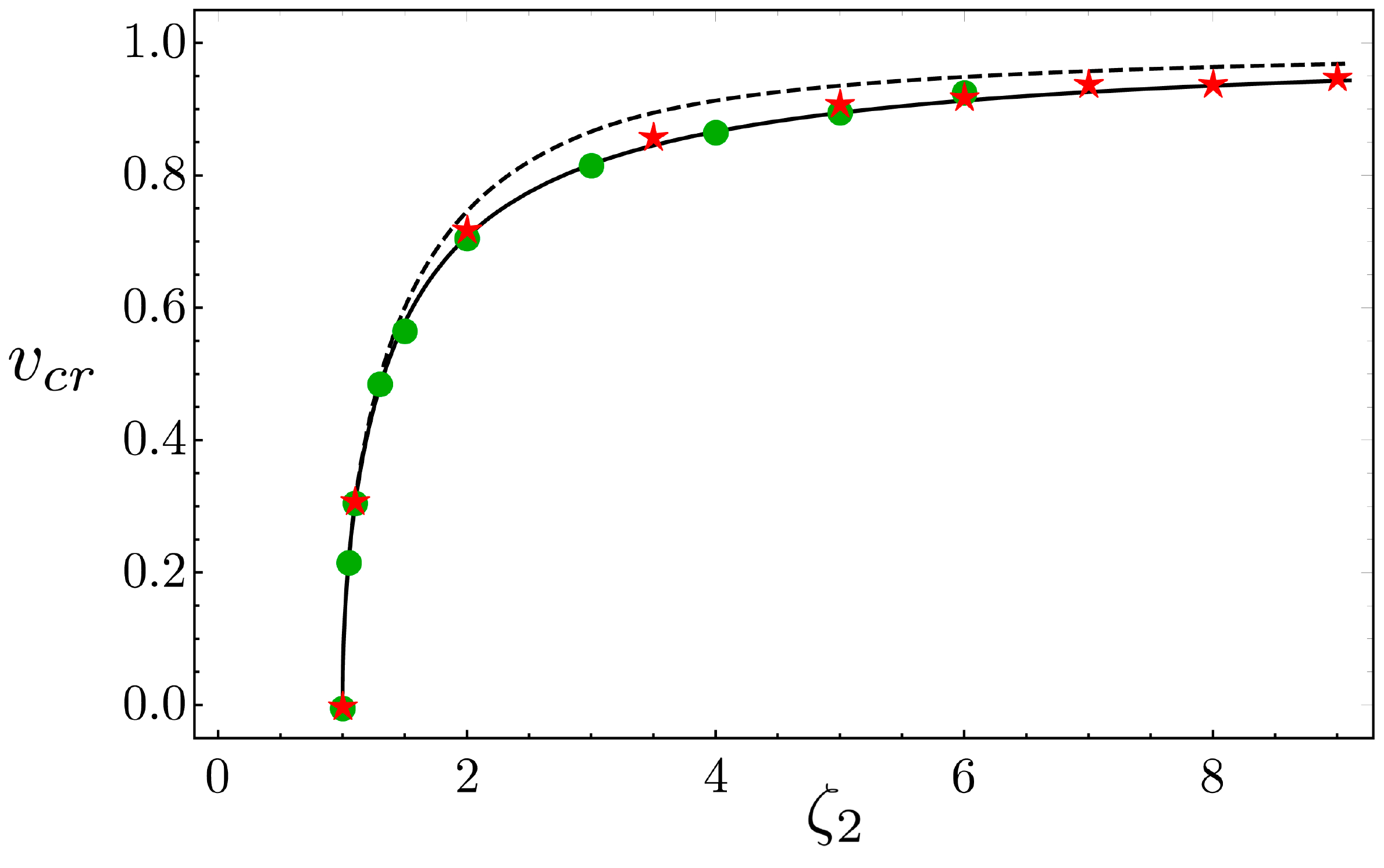}}
\caption{\label{fig:case1_vcr}
The critical velocity $v_{cr}$ dependence on the coupling constant jump ($\zeta_1=1$). 
The solid line shows theoretical curve Eq.~\eqref{eq:vcr_theor}, red stars 
and green circles represent the numerical results for the sine-Gordon and the $\varphi^4$ models, respectively. The dashed line shows results from the perturbation theory see (Eq.~(S25) in \cite{sm}).}
\end{figure}

To describe this phenomenon quantitatively we use the conservation of the energy of the system
\begin{equation}\label{eq:ener_generalform}
	E[\varphi]=\int\limits_{-\infty}^{+\infty}\left[\frac{1}{2}\varphi_t^2+\frac{1}{2}\varphi_x^2+\Theta(t,x)\, U(\varphi)\right]dx,
\end{equation}
and assume that the contribution of the radiation modes is negligible. So taking into account Eq.~\eqref{eq:sol_generalform} we can approximate with the accuracy $O(e^{-|x_0\sqrt{\zeta_1}|})$ energy in the initial moment of time as
\begin{equation}
E^{(0)} = \frac{\sqrt{\zeta_1}M}{\sqrt{1-v_{in}^2}},\enskip M = \int\limits_{-\infty}^\infty dx \left[
\frac{1}{2} F_x^2+ U(F(x))
\right].
\end{equation}
This is nothing but an energy of a relativistic particle moving with velocity $v_{in}$. When the solitary wave is completely transmitted to the area with the coupling constant $\zeta_2$, the energy evaluated with the same accuracy reads 
\begin{equation}
E_{2} = \frac{\sqrt{\zeta_2}M}{\sqrt{1-v_f^2}}.
\end{equation}
Imposing energy conservation in this form we obtain 
\begin{equation}\label{eq:vf_theor}
v_{f} =\sqrt{\frac{\zeta_2}{\zeta_1}(v_{in}^2-1)+1}.
\end{equation}
We see that Eq.~\eqref{eq:vf_theor} always has solutions for $\zeta_1>\zeta_2$, meaning that the wave is always transmitted to the less ``dense" domain. For $\zeta_1<\zeta_2$ the wave will be completely reflected if the initial velocity $v_{in}$ is lower than the critical value
\begin{equation}\label{eq:vcr_theor}
v_{cr} =\sqrt{1-\frac{\zeta_1}{\zeta_2}}.
\end{equation}
This formula has already appeared in Ref.~\cite{sasha2} (see Eq.~(5.23) there) and later in Ref.~\cite{ff} and book \cite{ff2} with a typo that we correct in this manuscript. 
We check numerically for both models how the critical velocity depends on the jump of the coupling constants and find remarkable coincidence with the theoretical prediction Eq.~\eqref{eq:vcr_theor} (see Fig.~\ref{fig:case1_vcr}). Notice that contrary to \cite{sasha2,ff,ff2} the agreement holds even in the non-perturbative regime. We also check predictions of the perturbation theory \cite{sm}. The corresponding critical velocity is shown as a dashed line in Fig.~\ref{fig:case1_vcr}. We see that, surprisingly, even for the large perturbations $\zeta_2\gg \zeta_1$ the perturbation theory 
results are qualitatively correct, and quantitatively applicable for $\zeta_2\sim \zeta_1$.

\begin{figure}[h]
\center{\includegraphics[width=0.9\linewidth]{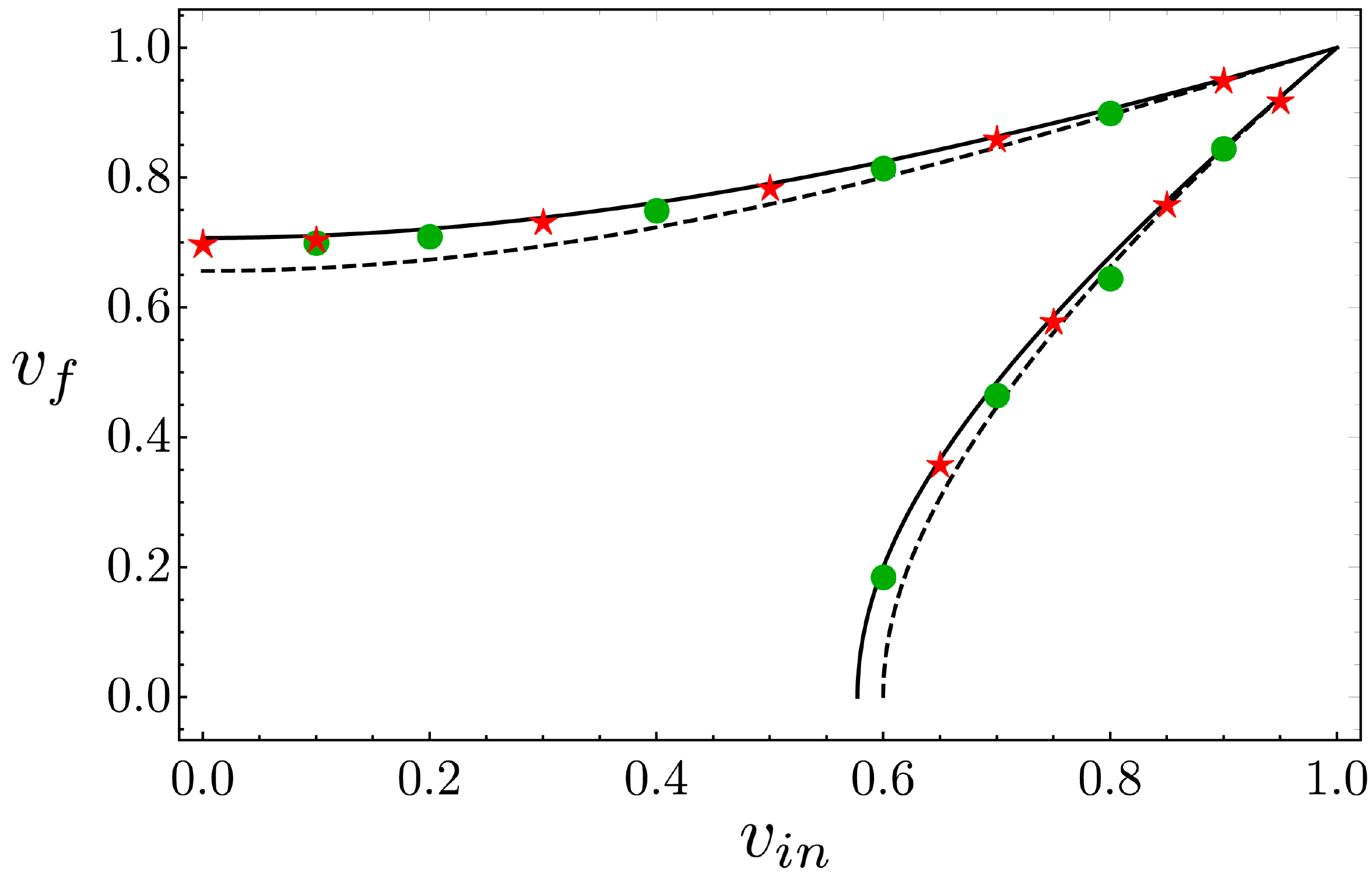}}
\caption{The final velocity $v_f$ as a function of the initial velocity $v_{in}$, for $\zeta_2/\zeta_1=0.5$ (upper set of points) and for $\zeta_2/\zeta_1=1.5$ the lower set of points. The solid lines show theoretical predictions Eq.~\eqref{eq:vf_theor}, red stars 
and green circles represent the numerical results for the sine-Gordon and the $\varphi^4$ models, respectively. The dashed line shows results from the perturbation theory Eq.~(S24) in \cite{sm}.}
\label{fig:case1_vf}
\end{figure}

Let us discuss the properties of the transmitted wave. For $\zeta_1>\zeta_2$, it immediately follows from Eq.~\eqref{eq:vf_theor} that ${v_f>v_{in}}$. Moreover, 
$v_f\neq 0$ even for $v_{in}=0$, which means that the static solitary wave placed at the boundary is ``pulled" into the less ``dense" area. The case $\zeta_2=0$ corresponds to the tunneling into a vacuum. In this case, the wave accelerates to the highest possible value $v_f=1$ regardless of the initial $v_{in}$. The profile keeps its soliton/kink-like form to assure the conservation of the topological charge 
\begin{equation}
    Q=\int_{-\infty}^{+\infty} \varphi_x\,dx = \varphi(+\infty)-\varphi(-\infty).
\end{equation}
The exact form can be found by taking limit $\zeta_2\to 0$ and $v_{in} \to 1$ in Eq.~\eqref{eq:sol_generalform}. In  Fig.~\ref{fig:case1_vf} we compare theoretical predictions for the final velocity for the fixed generic ratio $\zeta_1/\zeta_2$ for different $v_{in}$ with the results of numerical simulations. We see the perfect agreement, which confirms the validity of our assumptions and robustness of the particle structure.

For $\zeta_1<\zeta_2$ a small initial velocity does not allow the wave to 
be transmitted to the more ``dense'' area, resulting in the elastic reflection with a final velocity $v_f=-v_{in}$. The turning point lies close to the junction and depends on the initial velocity $v_{in}$. In some cases, the turning point  might be even in the more ``dense" part. For $v_{in}>v_{cr}$ the wave is completely transmitted and continues its propagation indefinitely.
We compare prediction by the theoretical formula Eq.~\eqref{eq:vf_theor} with the numerical simulations in Fig.~\ref{fig:case1_vf}. Contrary to the perfect match for the $v_{cr}$, we observe small deviations between theoretical and numerical 
value when the initial velocity is close to the critical. 
Indeed, the self-consistency of our approach fails as for $v_{in}= v_{cr}+0 $ the solitary wave should stick to the boundary as $v_{f}\approx 0$ forming some kind of the static profile. But in this case, the main assumption that the profile is situated on either side is violated and more refined analysis is needed. One can check analytically that there is no static solution with the prescribed boundary conditions and for $v_{in}= v_{cr}$ the soliton is reflected from the junction. Below we demonstrate what happens if the initially static soliton is placed exactly at the boundary between two media. 

\begin{figure}[h]
\center{\includegraphics[width=0.9\linewidth]{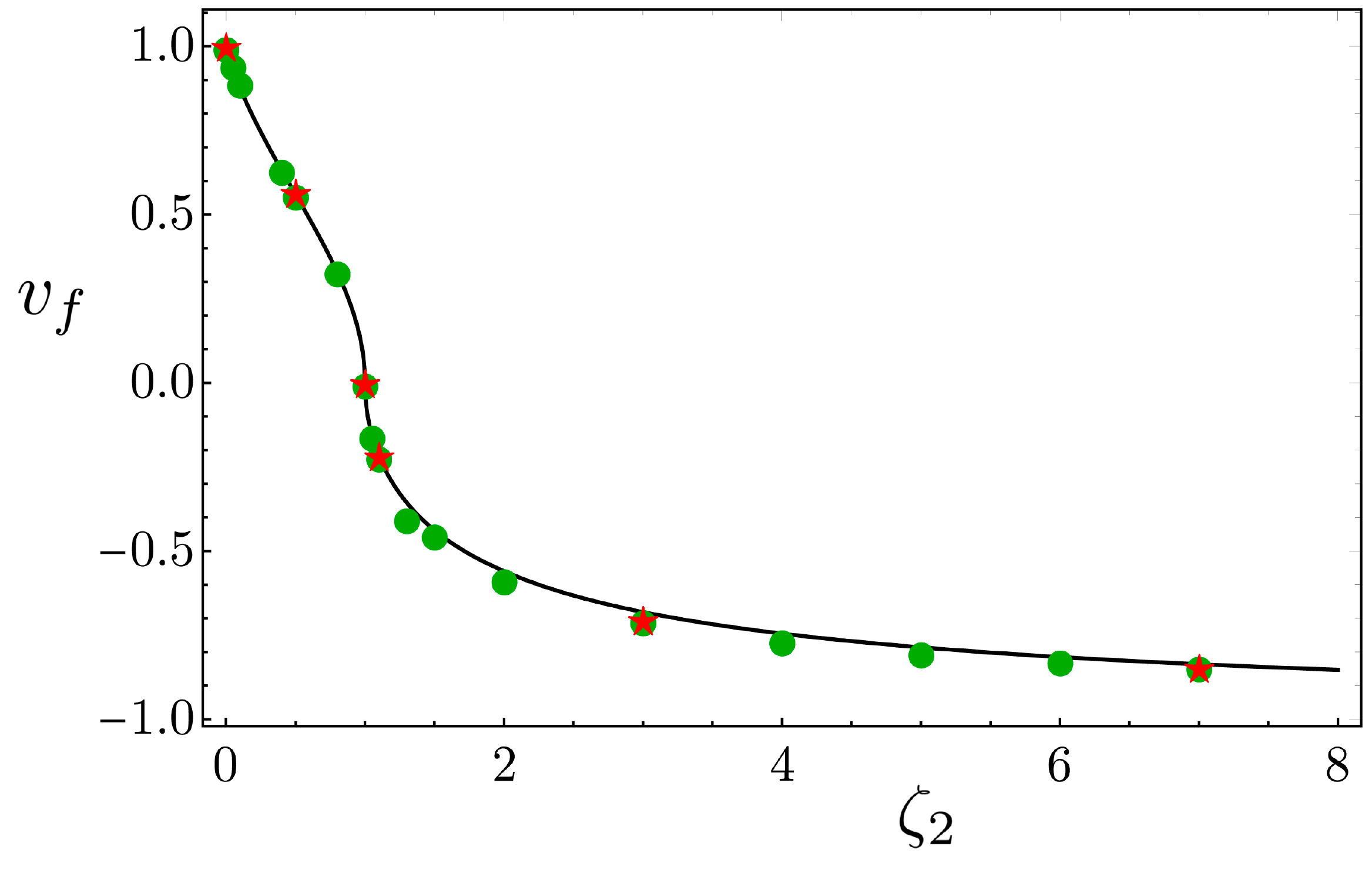}}
\caption{\label{fig:case2_vf} The final velocity $v_f$ for the initially static profile located in boundary of $\zeta_1=1$ and various $\zeta_2$. 
The solid line shows theoretical prediction Eq.~\eqref{eq:case2a_theory}, red stars 
and green circles represent the numerical results for the sine-Gordon and the $\varphi^4$ models, respectively.}
\end{figure}

{\it Wave on the border.} We consider the initial profile consisting 
of two static solutions ($v_{in}=0$) in each part of the system glued together at the junction. For the $\zeta_1\neq \zeta_2$ the profile starts moving to the less ``dense" region, i.e. with lower potential energy, while keeping the form of a kink. Namely, for $\zeta_1>\zeta_2$ the profile will move to the right part and for $\zeta_1<\zeta_2$ to the left, in contrast to~\cite{diode3}, when for similar setup the wave would stick near the junction point. Again, assuming that to satisfy the topological constraint the profile in the uniform part will take the form given by Eq.~\eqref{eq:sol_generalform} we can predict its velocity from the energy conservation law
\begin{equation}
   \frac{M}{2}\left(\sqrt{\zeta_1}+\sqrt{\zeta_2}\right)=\frac{M\sqrt{\text{min}\,[\zeta_1,\zeta_2]}}{\sqrt{1-v_f^2}}.
\end{equation}
The sign for $v_f$ has to be chosen such that the wave moves in the less ``dense" area, then the final velocity can be presented as
\begin{equation}\label{eq:case2a_theory}
    v_f = \text{sgn}[\zeta_1-\zeta_2]\,\displaystyle \sqrt{1-\frac{4\text{min}\,[\zeta_1,\zeta_2]}{(\sqrt{\zeta_1}+\sqrt{\zeta_2})^2}}.
\end{equation}
This result again turns out to be model-independent. Comparison of the numerically obtained $v_{f}$ with the theoretical prediction in Fig.~\ref{fig:case2_vf} shows the perfect agreement. Note that the existence of a force pushing the fluxons in Josephson junction in the direction of smaller self-energy in the presence of a temperature gradient can be seen in experiments~\cite{diode1}.

\begin{figure}[h]
\center{\includegraphics[width=0.90\linewidth]{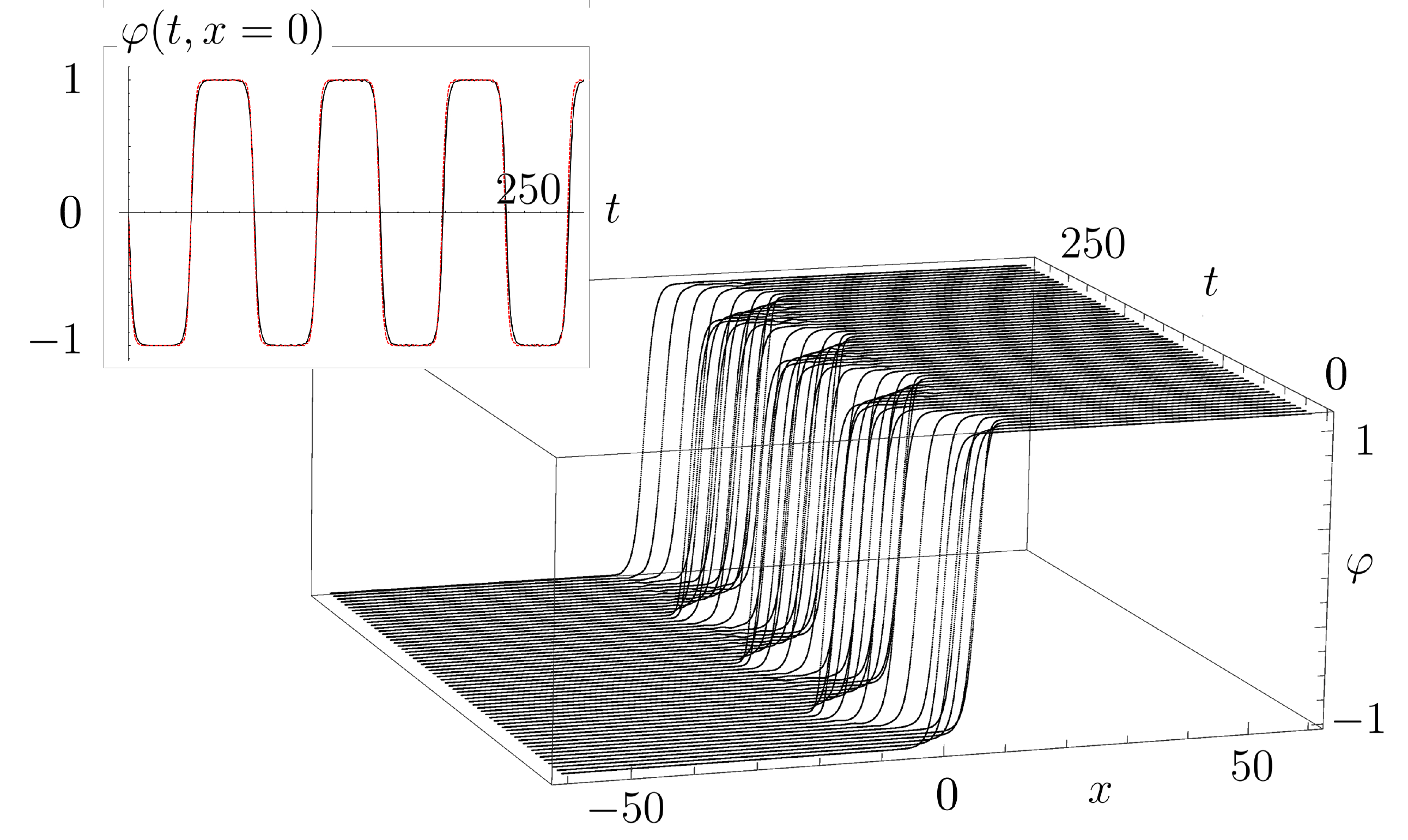}}
\caption{The solitary wave locked in the ``sandwich'' structure with the inner  section ($10<|x|$) corresponding to $\zeta_1=1$ and the outer sections $|x|>10$ with $\zeta_2=5$. The numerical simulations are done for the $\varphi^4$ model. The inset shows time-dependence for the field in the middle ($x=0$). The black line shows numerical results for the $\varphi^4$ model and the red dashed line stands for the scaled ($\varphi(t,x=0)/(2\pi)$) data for the  sine-Gordon model.}
\label{fig:case2a}
\end{figure}

{\it Locked wave.} By designing specific heterostructures one can achieve even more exotic behavior of a solitary wave. For instance, let us consider a ``sandwich" structure with two junctions dividing the space into three regions, that are described by the modified Eq.~\eqref{eq:theta1} as follows
\begin{equation}
    \Theta^{I}(t,x)=\zeta_2\theta(|x|-b)+
    \zeta_1\theta(b-|x|),\enskip-\infty<t<\infty.
\end{equation}
Here, $b>0$ defines the boundaries of the sections and is assumed to be much larger than the width of solitary waves. We set the initial profile in the middle section in a form of the soliton or the kink moving with the subcritical velocity $0<v_{in}<v_{cr}$  given in Eq.~\eqref{eq:vcr_theor}. For such a choice of $v_{in}$ the solitary wave would reflect from the boundaries at $x=\pm b$ unable to escape the middle region and moves freely inside similar to a photon between two mirrors. We check this behavior for both models (see Fig.~\ref{fig:case2a}) and observe that the form and the velocity of the profile remain unchanged even after multiple reflections, which proves the negligent role of the radiation modes. 
Note also that for the layered structures for NLS systems there are static solutions for certain parameters \cite{PhysRevE.73.066619,PhysRevA.87.043812,PhysRevA.87.063849}. 
Similar solutions also exist for the piece-wise time-dependence for Toda chain and NLS \cite{KOMINIS_2010,PhysRevE.88.042924,PhysRevE.81.066601}. 
We address time dependent for our systems below.

{\it Quench.} We consider the protocol given by Eq.~\eqref{eq:theta2} that corresponds to the instant change of coupling constant from $\zeta_1=1$ to some finite value $\zeta_2>0$. This quench is equivalent to the re-scaling of $t$ and $x$, similar to Ref. \cite{Gamayun_2016}. As the initial condition we consider a solitary wave Eqs.~\eqref{eq:sol_generalform} moving with the initial velocity $v_{in}>0$.
The dynamics of the sine-Gordon model after the quench is still integrable and can be addressed by means of the inverse scattering transformation technique (IST) \cite{Faddeev_1987} (see the Supplementary Materials \cite{sm}). 
Even though this method is an exact one, the corresponding computations are technically involved and the closed results can be obtained only for the certain initial profiles. Namely, we will find that if there exists a positive integer $n$, such that 
\begin{equation}\label{param}
    (2n+1)^2 =v_{in}^2 + \zeta_2 (1 - v_{in}^2 )
\end{equation}
then the resulting profile will consist of the soliton moving with the velocity 
\begin{equation}\label{qv}
    v_f = \frac{v_{in}}{2n+1}
\end{equation}
and $n$ breathers with the zero velocity.  It is interesting to notice that 
for the similar quench for the $\varphi^4$ model no breathers appear, but only the regular moving kink.

\begin{figure}[t]
\center{\includegraphics[width=0.9\linewidth]{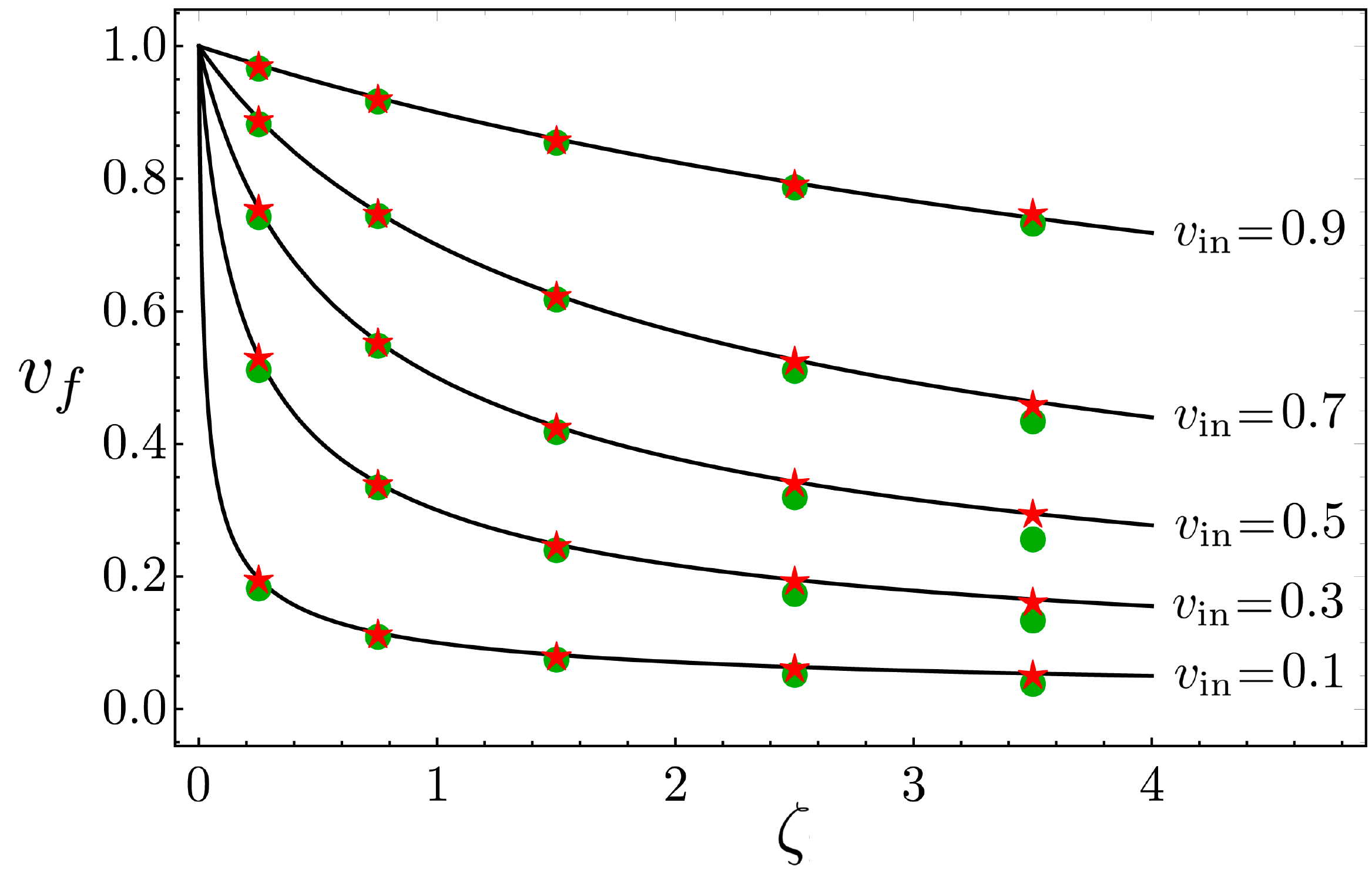}}
\caption{\label{fig:case3_vf} The final velocity $v_f$ and the function of the quench parameter $\zeta$ for different initial velocities.  The solid lines show theoretical prediction Eq.~\eqref{eq:case3_vf}, red stars 
and green circles represent the numerical results for the sine-Gordon and the $\varphi^4$ models, respectively.}
\end{figure}

To describe a generic $\zeta_2$ and address the $\varphi^4$ model we proceed analogously to the spatially inhomogeneous case. We notice that the total momentum of the field 
\begin{equation}
    P=\int_{-\infty}^{+\infty}\varphi_t\,\varphi_x\,dx
\end{equation}
is conserved in both models. Assuming that the total value of the momentum is saturated mainly by the soliton (kink) i.e. neglecting the radiation modes, and using the profile from Eq.~\eqref{eq:sol_generalform}, we obtain 
\begin{equation}
    P_{1,2} = \frac{-\sqrt{\zeta_{1,2}}\, v_{1,2}}{\sqrt{1-v_{1,2}^2}}\int_{-\infty}^{+\infty} F^{\prime}(X)\,dX,
\end{equation}
where $P_1$ and $P_2$ are total momenta before and after the quench, respectively. 
From the conservation law $P_1 = P_2$ we immediately obtain model independent prediction for the final velocity ($\zeta_1=1$)
\begin{equation}\label{eq:case3_vf}
    v_f=\frac{v_{in}}{\sqrt{v_{in}^2+\zeta_2(1-v_{in}^2)}}.
\end{equation}
This result coincides with the one obtained from the IST given by Eq.~\eqref{qv} for a particular value of the parameters in Eq.~\eqref{param}. For generic values $\zeta_2$ we compare these predictions with the numerical results in Fig.~\ref{fig:case3_vf} and see the perfect match.

\begin{figure}[t]
\center{\includegraphics[width=0.99\linewidth]{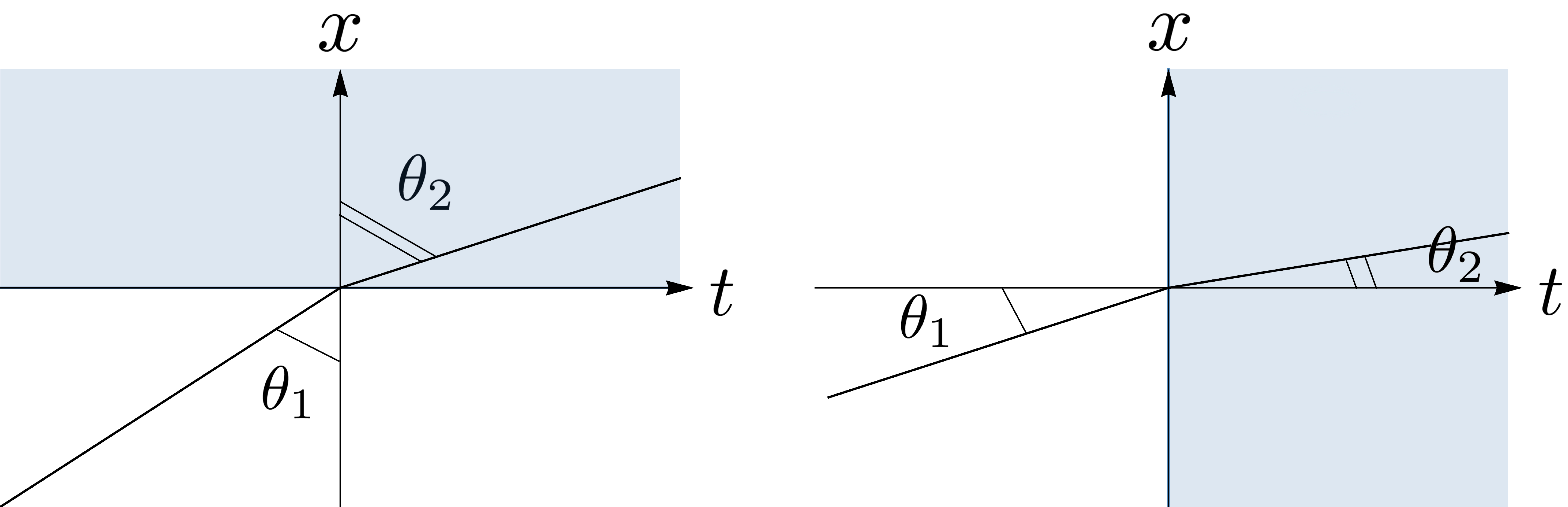}}
\caption{The schematic presentation for the center of the solitary wave trajectory. The shaded parts show the regions with the altered coupling constant. The left panel corresponds to the junction and the right to the quench protocol.}
\label{figSnell}
\end{figure}

{\it The hyperbolic Snell's law.} It is interesting to note that the soliton's center trajectory plotted in $(x,t)$ plane reminds the light refraction between the two media. The velocity is related to the angle $\theta$ measured from the normal as $v=\cot \theta$ for the spatial junction (see Fig.~\ref{figSnell}). The corresponding range of angles is $\theta \in [\pi/4,\pi/2]$. The temporal inhomogeneity (quench) covers the rest of angles $\theta \in [0,\pi/4]$; the relation to the velocity reads as $v=\tan \theta$. Further, we may identify the refractive index of the material as $n=\sqrt{\zeta}$. The traditional Snell's law
\begin{equation}
    n_1 \sin \theta_1 = n_2 \sin \theta_2
\end{equation}
would imply the following relations 
\begin{equation}
    \frac{\zeta_1}{1+v_1^2} = \frac{\zeta_2}{1+v_2^2},\qquad 
    \frac{\zeta_1v^2_1}{1+v_1^2} = \frac{\zeta_2v^2_2}{1+v_2^2}
\end{equation}
for the junction and quench protocols, respectively. To get an observed 
final velocities Eqs.~\eqref{eq:vf_theor} and \eqref{eq:case3_vf} from the Snell's law, one has to replace $v_1^2 \to - v_{in}^2$ and $v_2^2 \to - v_f^2$, which is consistent with the Wick rotation $t \to -i\tau$ that renders the time as a genuine spatial coordinate.

{\it Conclusion and outlook.} 
To conclude, we have studied the propagation of solitary waves in the sine-Gordon and $\varphi^4$ field theories in the presence of spatial inhomogeneity in the form of a junction, and under a quench protocol. We have argued that solitary waves remain robust under these non-equilibrium conditions. This fact allowed us to describe their dynamics in a universal way.  In particular, we  have demonstrated that the relation between the final and the initial velocities of the solitary wave can be cast in the form of the hyperbolic analog of the Snell's law. We have corroborated our general results by model-specific exact calculations in the framework of the  inverse scattering approach, by extensive numerical simulations and by perturbative calculations. Interestingly, the predictions of the perturbation theory appear to be qualitatively correct even beyond its formal range of applicability.

Let us emphasize that the robustness is guaranteed by the 
topological nature of the waves, and the ``quantized'' value of the 
overall phase. For comparison, in the NLS equation, the overall phase is conserved but can take arbitrary values, so the radiation can accumulate the finite phase difference over the system, leading to the destruction of the soliton ~\cite{PhysRevA.91.031605,PhysRevA.99.023605}. 

Perhaps, the easiest way to experimentally see the presented phenomena is in the non-uniform Josephson junctions described in terms of the non-uniform sine-Gordon model~\cite{Remoissenet,ustinov}. Earlier, the motion of solitary waves in the Josephson junction that can be controlled or modulated by a magnetic field was discussed in~\cite{diode1,diode2}, both for experimental and theoretical setups. All the mentioned results are looking quite promising for implementing in telecommunications or rapid single flux quantum logic through a fluxon diode.

{\it Acknowledgments.} We are grateful to Oleg Lychkovskiy for careful reading of the manuscript and many fruitful remarks. O. G. acknowledges the support from the European Research Council under ERC Advanced grant 743032 DYNAMINT. This work is part of the Delta Institute for Theoretical Physics (DITP) consortium, a program of the Netherlands Organization for Scientific Research (NWO) that is funded by the Dutch Ministry of Education, Culture and Science (OCW).

\bibliography{bibliography.bib}

\end{document}


\title{Supplemental Material for ``Solitary wave propagation in media with step-like inhomogeneities''}
	
	\author{Mariya Lizunova}
	\affiliation{Institute for Theoretical Physics, Utrecht University, Princetonplein 5, 3584 CC Utrecht, The Netherlands}
	\affiliation{Institute for Theoretical Physics Amsterdam, University of Amsterdam, Science Park 904, 1098 XH Amsterdam, The Netherlands}
	
	\author{Oleksandr Gamayun}
	\email{o.gamayun@uva.nl}
	\affiliation{Institute for Theoretical Physics Amsterdam, University of Amsterdam, Science Park 904, 1098 XH Amsterdam, The Netherlands}
	\affiliation{Bogolyubov Institute for Theoretical Physics
		14-b Metrolohichna str. Kyiv, 03143, Ukraine}

	\date{\today}
	
	\maketitle
	
	\section{S1. Sine-Gordon inverse scattering transformation approach}
	
	Let us briefly review the inverse scattering transformation approach (IST) for the $(1+1)-$dimensional sine-Gordon equation. We follow mainly Ref. \cite{Faddeev_1987a}. The sine-Gordon equation
	\begin{equation}
		\varphi_{tt} - \varphi_{xx} + \zeta \sin \varphi=0
	\end{equation}
	is equivalent to the zero curvature condition  
	\begin{equation}
		U_t - V_x + [U,V] =0,
	\end{equation}
	where
	\begin{equation}\label{alp3}
		U =\frac{1}{4i}
		\left(
		\varphi_t \sigma_3  +\sqrt{\zeta} (\lambda+1/\lambda) \sigma_1\sin \frac{\varphi }{2}
		+ \sqrt{\zeta}(\lambda-1/\lambda) \sigma_2\cos \frac{\varphi }{2}\right),\end{equation}
	\begin{equation}
		V= \frac{1}{4i}
		\left(
		\varphi_x \sigma_3  + \sqrt{\zeta}(\lambda-1/\lambda) \sigma_1\sin \frac{\varphi }{2}
		+ \sqrt{\zeta}(\lambda+1/\lambda) \sigma_2\cos \frac{\varphi }{2}\right),
	\end{equation}
	where $\sigma_1$, $\sigma_2$ and $\sigma_3$ are the standard Pauli matrices and $\lambda$ is an arbitrary complex number called the spectral parameter. Further, we focus mainly on $\zeta =1$. 
	The Jost solutions $T_{\pm}(x)$ are defined as the solutions of the auxiliary linear problem 
	\begin{equation}\label{alp22}
		\frac{d T_{\pm}(x)}{dx} = U T_{\pm}(x),
	\end{equation}
	specified by the following asymptotic behavior
	\begin{equation}
		T_{-}(x\to -\infty) = \mathcal{E} e^{-ix \sigma_3 (\lambda^2-1)/(4\lambda)},\qquad 
		T_+(x\to +\infty) = \begin{cases}
			(-1)^{Q/2} \mathcal{E} e^{-ix \sigma_3 (\lambda^2-1)/(4\lambda)}, &  Q  \quad {\rm even,}  \\
			(-1)^{(Q-1)/2} i\sigma_3\mathcal{E} e^{-ix \sigma_3 (\lambda^2-1)/(4\lambda)},& Q  \quad {\rm odd,}
		\end{cases}
	\end{equation}
	where the matrix $\mathcal{E}$ and the topological charge $Q$ are given by
	\begin{equation}
		\mathcal{E} = \frac{1}{\sqrt{2}} \left(
		\begin{array}{cc}
			1 & i \\
			i & 1
		\end{array}
		\right),\qquad Q =   \frac{1}{2\pi} \int\limits_{-\infty}^\infty dx \varphi_x(x).
	\end{equation}
	The Jost solutions are related by the transfer matrix $T(\lambda)$, whose elements are called the scattering data 
	\begin{equation}\label{tt}
		T_- (x,\lambda) = T_+(x,\lambda) T(\lambda),\qquad T(\lambda) = \left(
		\begin{array}{cc}
			a(\lambda) & - b^*(\lambda) \\
			b(\lambda) & a^*(\lambda)
		\end{array}
		\right).
	\end{equation}
	The evolution given by the sine-Gordon equation looks extremely simple on the scattering data
	\begin{equation}\label{dyn1}
		a(\lambda,t) = a(\lambda,0),\qquad b(\lambda,t)= b(\lambda,0) e^{it(\lambda^2+1)/(2\lambda) }.
	\end{equation}
	Additionally, the scattering data should be supplemented by the discrete spectrum that consists of the zero of $a(\lambda)$ in the upper half complex plane. The solitons correspond to the purely imaginary zeroes $a(i\ae_k)=0$, $\ae_k>0$ and the breathers to the symmetric pairs of zeroes $\lambda_k$ and $-\lambda^*_k$, with ${\rm Im}\lambda_k, {\rm Re}\lambda_k >0$.
	The associated scattering data is the proportionality coefficient between the first column of $T_-$ and the second column of $T_+$, namely
	\begin{equation}
		T_-^{(1)}(x,\lambda_k) = \gamma_k T_{+}^{(2)}(x,\lambda_k),\qquad k = 1, \dots ,n.
	\end{equation}
	Here, $n$ is a number of zeros and the upper indices indicate the corresponding columns in the matrices. 
	
	The dynamics of the proportionality coefficient is given by 
	\begin{equation}\label{dyn2}
		\gamma_k(t) = e^{it(\lambda_k^2+1)/(2\lambda_k) } \gamma_j(0).
	\end{equation}
	For instance, the one-soliton solution from the dynamical variables can be presented as 
	\begin{equation}\label{11sol}
		\varphi(x,t) = -4 \arctan \frac{e^{x(\ae^2+1)/(2\ae)}}{\gamma(t)} = 4 \arctan \text{exp}\left(\frac{x- v t - x_0}{\sqrt{1-v^2}}\right),
	\end{equation}
	where parameters $v$ (defined as $v_{in}$ in the main text) and $x_0$ are related to $\ae$ and $\gamma(0)$ as
	\begin{equation}
		v = \frac{1-\ae^2}{1+\ae^2}, \qquad -e^{x_0/\sqrt{1-v^2}}  = \gamma(0).
	\end{equation}
	The corresponding Jost solutions are 
	\begin{equation}\label{Jost}
		T_+ = \frac{\mathcal{E}}{\sqrt{1+e^{2\xi}}}\left(\begin{array}{cc}
			\frac{\lambda+ i \ae}{\lambda - i \ae}  & -e^{\xi} \\
			e^{\xi}& \frac{\lambda- i \ae}{\lambda + i \ae} 
		\end{array}\right)e^{-i\sigma_3 x \frac{\lambda^2-1}{4\lambda}}
		, \qquad
		T_- = \frac{\mathcal{E}}{\sqrt{1+e^{2\xi}}}\left(\begin{array}{cc}
			1 & -\frac{\lambda+ i \ae}{\lambda - i \ae}e^{\xi} \\
			\frac{\lambda- i \ae}{\lambda + i \ae}e^{\xi}&  1
		\end{array}\right)e^{-i\sigma_3 x\frac{\lambda^2-1}{4\lambda}},
	\end{equation}
	where $\xi = (x- x_0 -v t)/\sqrt{1-v^2}$, which means $e^{\xi} = -e^{x(1+\ae^2)/(2\ae)}/\gamma(t)$. 
	The transfer matrix $T(\lambda)$ is diagonal for this case (see Eq.~\eqref{tt})
	\begin{equation}
		a(\lambda) = \frac{\lambda- i \ae }{\lambda+ i \ae },\qquad b(\lambda) = 0.
	\end{equation}
	
	\section{S2. Perturbation theory}
	
	In this section, we consider the perturbation theory based on the IST \cite{Kaupa,Karpman2a} (for a general review see Ref. \cite{RevModPhys.61.763a}). The sine-Gordon equation with the spatial inhomogeneity reads as 
	\begin{equation}
		\varphi_{tt} - \varphi_{xx} + \sin \varphi = - (\zeta -1)\theta(x) \sin \varphi  = - i \delta  \theta(x) \sin \varphi.
	\end{equation}
	Here, $\theta(x)$ is the Heaviside step function. 
	If we choose $U$ and $V$ as in Eq.~\eqref{alp3} then the zero curvature condition will transform into 
	\begin{equation}
		U_t - V_x + [U,V] = - \frac{ (\zeta -1)}{4i} \theta(x) \sigma_3 \sin \varphi  \equiv \mathcal{P}[\varphi].
	\end{equation}
	The continuous and discrete scattering data evolves as 
	\begin{equation}\label{pert1}
		T_t(\lambda) - i\frac{\lambda^2+1}{4\lambda} [\sigma_z ,T(\lambda)] =\int\limits_{-\infty}^{\infty} dz T_+^{-1}(z)\mathcal{P}[\varphi(z)]T_-(z),
	\end{equation}
	\begin{equation}\label{pert2}
		\frac{d\gamma}{dt}- \frac{1-\ae^2}{2\ae } \gamma = \frac{i}{\dot{a}(i\ae)}\int\limits_{-\infty}^{\infty} dz  \left[\dot{T}^{(1)}_-(z) -\gamma\dot{T}^{(2)}_+(z) \right]^{T}\sigma_2\mathcal{P}[\varphi(z)]T^{(1)}_-(z), 
	\end{equation}
	\begin{equation}\label{pert3}
		i\frac{d\ae}{dt}  =  \frac{i}{\dot{a}(i\ae)}\int\limits_{-\infty}^{\infty} dz  \left[T_+^{(2)}(z)\right]^{T}\sigma_2\mathcal{P}[\varphi(z)]T^{(1)}_-(z).
	\end{equation}
	The LHS of Eqs. \eqref{pert2} and \eqref{pert3} should be evaluated at $\lambda  = i \ae$. Replacing transfer matrix by its one-soliton value given by Eq.~\eqref{Jost} yields
	\begin{equation}
		\frac{d \ae}{dt} = \frac{2 (\zeta -1) \ae^2}{1+\ae^2} \frac{\gamma^2}{(1+\gamma^2)^2},
	\end{equation}
	\begin{equation}
		\frac{d \gamma}{dt} = \frac{1-\ae^2}{2\ae } \gamma  \left(1- 2 (\zeta -1)   \frac{\ae^2}{(1+\ae)^2} \frac{\gamma^2}{1+\gamma^2} \right).
	\end{equation}
	This dynamical system has an integral of motion
	\begin{equation}
		I(t) = \frac{1+\ae^2}{2\ae} +  (\zeta -1)\frac{ \ae}{1+\ae^2} \frac{\gamma^2}{1+\gamma^2},\qquad \frac{d I(t)}{dt} =0.
	\end{equation}
	At the initial moment, the soliton is localized far from the border, which means that $\gamma(0)\approx 0$, if the soliton is transmitted then $\gamma(+\infty)\to -\infty$. From this, we can determine the finite velocity
	\begin{equation}
		v_f = \frac{ 2 \sqrt{1-2  (\zeta -1)(1-v^2)}}{ \sqrt{1-2  (\zeta -1)(1-v^2)}+1}\sqrt{1-\frac{4 \left(1-v^2\right)}{\left( \sqrt{1-2  (\zeta -1)(1-v^2)}+1\right)^2}},
	\end{equation}
	The critical velocity can be determined from the condition that $v_f$ is real, which for $\zeta > 1$  reads as
	\begin{equation}
		v_{cr} = \frac{\sqrt{(\zeta -1)(\zeta+3)}}{\zeta+1}\theta(\zeta <3) + \theta(\zeta>3) \sqrt{\frac{2\zeta-3}{2(\zeta-1)}}.
	\end{equation}
	
	\section{S3. Quench}
	
\begin{figure}[h]
		\center{\includegraphics[width=0.5\linewidth]{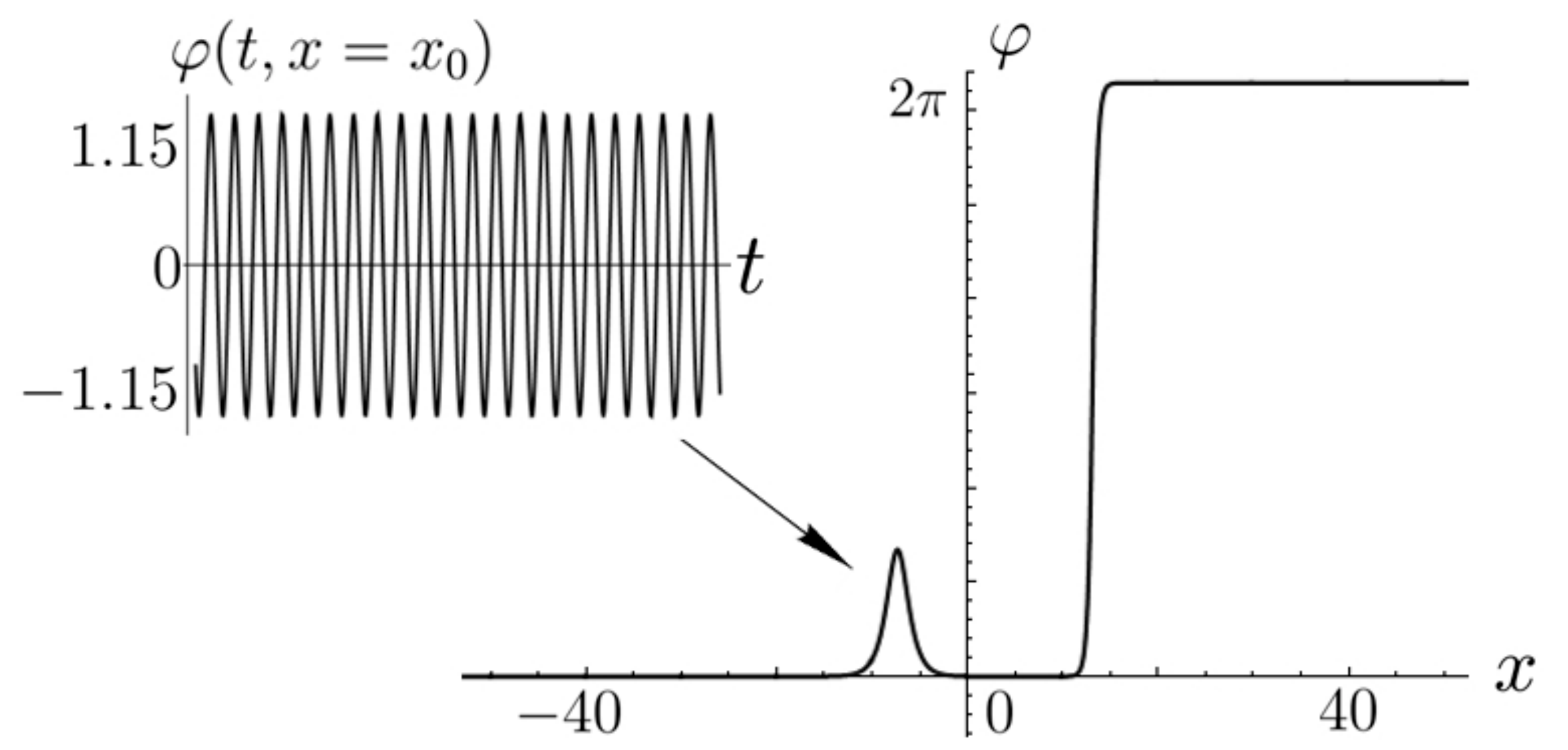}}
		\caption{The numerically obtained snapshot of the post quench profile for initial velocity $v=0.2$ and $\zeta = (9-v^2)/(1-v^2)$. The inset shows oscillations in the initial point of the soliton center.}
		\label{fig}
	\end{figure}

	In this section, we compute scattering data that corresponds to the quench protocol. This requires solving linear system Eq.~\eqref{alp22} with the connection given by Eq.~\eqref{alp3}
	with arbitrary $\zeta$, evaluated on  $\zeta=1$ one-soliton solution Eq.~\eqref{11sol}. We set $t=0$, $x_0=0$ and introduce variables
	\begin{equation}
		y=\frac{x}{\sqrt{1-v^2}}, \qquad k_\lambda = \frac{\sqrt{\zeta}\sqrt{1-v^2}}{4\lambda} (\lambda^2-1),\qquad \omega_\lambda = \frac{\sqrt{\zeta}\sqrt{1-v^2}}{4\lambda} (\lambda^2+1).
	\end{equation}
	The linear problem then reads as 
	\begin{equation}
		\cosh(y)\frac{dT}{dy} = i\left(
		\frac{ v\sigma_3}{2}  -\omega_\lambda\sigma_1+k_\lambda \sinh(y)\sigma_2\right)T,
	\end{equation}
	\begin{equation}
		\cosh(y)\frac{dT}{dy} = \frac{i}{2}\left(
		v \sigma_3 -\frac{\sqrt{\zeta}\sqrt{1-v^2}}{2\lambda}(\lambda^2+1)\sigma_1+\frac{\sqrt{\zeta}\sqrt{1-v^2}}{2\lambda} (\lambda^2-1)\sinh(y)\sigma_2\right)T.
	\end{equation}
	We perform the following change of variables
	\begin{equation}
		z = \frac{1+ \tanh (y)}{2},\qquad T = \frac{\mathcal{E}}{\cosh (y)^{ik_\lambda}} \left( 
		\begin{array}{cc}
			0 & e^y \\ 1 & 0
		\end{array}
		\right) \psi
	\end{equation}
	to present the linear system in the form
	\begin{equation}
		\frac{d \psi}{dz} = \left(\frac{A_0}{z} + \frac{A_1}{1-z} \right)\psi 
	\end{equation}
	with 
	\begin{equation}
		A_0  = -\frac{1}{2}\left(
		\begin{array}{cc}
			0 & 0 \\
			v+2 i \omega _{\lambda } & 1+2 i k_{\lambda } \\
		\end{array}
		\right) ,\qquad A_1 = \frac{1}{2}\left(
		\begin{array}{cc}
			0 & v-2i \omega _{\lambda } \\
			0 & 2 i k_{\lambda }-1 \\
		\end{array}
		\right).
	\end{equation}
	These matrices are diagonalizable by the similarity transformation with matrices $G_i$, i.e. 
	$A_i = G_i {\rm diag}(0,\lambda_i) G_i^{-1}$. Therefore, the asymptotic behavior of $\psi$ is fixed up to constant matrices $C_i$ as
	\begin{equation}
		\psi(x\to-\infty) = G_0 \left(
		\begin{array}{cc}
			1 & 0 \\
			0 & e^{-y-2 i k_{\lambda }y} \\
		\end{array}
		\right)C_0,\qquad 
		\psi(x\to+\infty) = G_1 \left(
		\begin{array}{cc}
			1 & 0 \\
			0 & e^{-y+2 i k_{\lambda }y} \\
		\end{array}
		\right)C_1.
	\end{equation}
	Comparing with Eq.~\eqref{Jost} one obtains $C_0$ and $C_1$ and using 
	Ref. \cite{Gamayun_2016a} we recover the scattering data from Eq.~\eqref{tt} 
	\begin{equation}
		a(\lambda) = \frac{(v/2-i\omega_\lambda)\Gamma(1/2-ik_\lambda)^2}{\Gamma(1-ik_\lambda+\Omega/2)\Gamma(1-ik_\lambda-\Omega/2)},\qquad
		b(\lambda) = -\frac{\cos \pi\Omega/2}{\sinh(\pi k_\lambda)}
	\end{equation}
	with 
	\begin{equation}
		\Omega = \sqrt{v^2+\zeta \left(1-v^2\right)}.
	\end{equation}
	Let us find zeros of $a(\lambda)$ in the upper half-plane. 
	For $0<\Omega<2$ the only zero is determined by the numerator
	\begin{equation}
		\lambda_0 = i \frac{\left(\Omega-v\right)}{\sqrt{\zeta} \sqrt{1-v^2}} = i \ae_0.
	\end{equation}
	It corresponds to a soliton with velocity
	\begin{equation}
		v_f = \frac{1-\ae_0^2}{1+\ae_0^2} = \frac{v}{\Omega}.
	\end{equation}
	For $\Omega=1$ the whole profile consists of this soliton. 
	For $\Omega >2$ additional zeroes appear due to the presence of Gamma functions in the denominator, which corresponds to additional breathers. 
	One can check that the velocities of these breathers are zero, thus, they do not contribute to the total momentum. Due to the symmetry $\lambda\to 1/\lambda$ the contribution of the radiation modes to the total momentum is also zero
	\begin{equation}
		P_r   = \frac{1}{\pi} \int\limits_0^\infty \ln(1-|b(\lambda)|^2) (1-1/\lambda^2) d\lambda = 0.
	\end{equation}
	For odd values of $\Omega$ the radiation is completely absent.  For instance, for $\Omega=3$ the quench parameter is $\zeta = (9-v^2)/(1-v^2)$ and the scattering data reads 
	\begin{equation}
		a(\lambda) =  \frac{\lambda - i \sqrt{(3-v)/(3+v)}}{\lambda + i \sqrt{(3-v)/(3+v)}}
		\frac{\lambda^2 - 2i \lambda/\sqrt{9-v^2}-1}{\lambda^2 + 2i \lambda/\sqrt{9-v^2}-1}
		, \qquad b(\lambda)=0.
	\end{equation}
	This corresponds to the soliton moving with the velocity $v_f = v/3$ and the breather with the zero velocity and the frequency $\omega_1 = \sqrt{8-v^2}/\sqrt{1-v^2}$, e.g., see Fig.~\ref{fig}.
	
	\providecommand{\noopsort}[1]{}\providecommand{\singleletter}[1]{#1}%